\documentclass[superscriptaddress,twocolumn,showpacs,amsmath,amssymb]{revtex4}
\usepackage[T1]{fontenc}
\usepackage{graphicx}
\usepackage{dcolumn}
\usepackage{bm}
\usepackage{array}
\renewcommand{\vec}[1]{\mbox{\boldmath $#1$}}

\begin{document}

\preprint{}

\title{
Sum rule approach to a soft dipole mode in $\Lambda$ hypernuclei}

\author{F. Minato}
\affiliation{
Nuclear Data Center, Japan Atomic Energy Agency, Tokai 319-1195, Japan}
\affiliation{
Institut d'Astronomie et d'Astrophysique, 
Universit\'{e} Libre de Bruxelles, Campus de la Plaine CP 226, 1050, 
Brussels, Belgium
}

\author{K. Hagino}
\affiliation{
Department of Physics, Tohoku University, Sendai 980-8578, Japan}
\date{\today}

\begin{abstract}
Applying the sum rule approach, we investigate
the energy of a soft dipole motion 
in $\Lambda$ hypernuclei, 
which results from a dipole oscillation of 
a $\Lambda$ hyperon against the core nucleus. 
To this end, we systematically study 
single-$\Lambda$ hypernuclei, from 
$^{16}_{\;\,\Lambda}$O to $^{208}_{\;\;\;\Lambda}$Pb, for which 
the ground state wave function is obtained 
in the framework of 
Hartree-Fock method with several Skyrme-type  
$\Lambda N$ interactions.
Our results indicate that 
the excitation energy of the soft dipole $\Lambda$ mode, 
$E_{sd\Lambda}$, decreases as the mass number increases. 
We find that the excitation energy is well parametrized as 
$E_{sd\Lambda}=26.6A^{-1/3}+11.2A^{-2/3}$ MeV as a function of 
mass number $A$. 
\end{abstract}
%
\maketitle
%
%
Hypernuclear physics has attracted lots of attention in recent 
years \cite{HT06}. 
A $\Lambda$ hyperon is 
free from the Pauli principle from nucleons, and thus 
can exist in a deep inside of a nucleus. 
Because of this property, a $\Lambda$ hyperon provides 
various kinds of impurity effect on the core nucleus.
One of the well-known examples is a shrinkage of 
intercluster distance 
in $^{7}_{\Lambda}$Li and $^{9}_{\Lambda}$Be 
\cite{lithium,beryllium,MBI83,HKMM99,HK11}.
Other examples include a disappearance of nuclear deformation 
due to a $\Lambda$ hyperon 
\cite{Myaing2008,Myaing2010,LZZ11,Isaka2011}, an influence on the 
triaxial degree of freedom \cite{Myaing2011,Isaka2012}, and an increase 
of fission barrier height \cite{MCH09}. 
In addition to the static properties, 
a $\Lambda$ hyperon can also alter nuclear dynamical motions.  
In particular, low-lying modes of excitation in 
$\Lambda$ hypernuclei have been studied with 
cluster model \cite{Hiyama2002},
ab-initio few-body calculations \cite{Nemura2002,Nemura2005}, 
shell model \cite{Gal2011}, 5-dimensional (5D) 
collective Bohr Hamiltonian \cite{Yao2011}, 
and anti-symmetrized 
molecular dynamics (AMD)\cite{Isaka2012}. 
Together with experimental data,
those theoretical studies would 
help us to understand not only the impurity effect on nuclear 
structure but also the characteristics of an effective 
$\Lambda N$ interaction.

In our previous study, we studied 
multipole vibrational motions of $\Lambda$ hypernuclei 
using random-phase approximation (RPA) \cite{Minato2012}. 
We have predicted a novel dipole mode, which 
we call the soft dipole $\Lambda$ mode, 
in a double-$\Lambda$ hypernucleus 
$^{\;18}_{\Lambda\Lambda}$O \cite{Minato2012}. 
This mode 
appears 
in the low energy region, and its strength is almost concentrated in a single peak 
with a magnitude of about a quarter of that of the giant 
dipole resonance (with the peak height in the strength distribution being 
about a half of that for the giant dipole resonance). 
A similar peak appears neither in 
other multipolarities nor in the dipole motion of 
the normal nucleus, $^{16}$O. 
From the transition density, we argued that 
this mode corresponds to 
a dipole oscillation of the two $\Lambda$ 
hyperons against the core nucleus.
Notice that this is similar 
to the soft dipole motion in neutron-rich halo nuclei \cite{Ikeda10}, 
in which weakly bound valence neutrons oscillate against the core nucleus.

In this paper, we 
systematically study the soft dipole $\Lambda$ mode 
from light to heavy hypernuclei. 
In order to estimate the excitation energy of the 
soft dipole $\Lambda$ mode, we employ the sum rule approach
\cite{SH90,STG92,KS95}. This approach provides a convenient way 
to estimate the excitation energy, as it can be evaluated only with 
the ground state wave function. This enables one to study the 
soft dipole $\Lambda$ mode in single-$\Lambda$ hypernuclei, 
whereas an application of RPA to 
signle-$\Lambda$ hypernuclei is much more complicated due to the 
broken time-reversal symmetry and half-integer spins. 

In the sum rule approach, the moments of the strength distribution 
are expressed in terms of the ground state expectation value of 
certain operators \cite{RS80,BLM79}. 
In order to apply this approach, 
we thus first obtain 
the wave function for the ground state 
of hypernuclei 
by adopting the Hartree-Fock method with a Skyrme-type zero-range force 
as an effective $\Lambda N$ interaction.
The $\Lambda N$ and 3-body $\Lambda NN$ interactions then read 
\cite{Rayet},
\begin{equation}
\begin{split}
&v_{\Lambda N}(\vec{r}_\Lambda-\vec{r}_N)
=
t_0^\Lambda(1+x_0^\Lambda P_\sigma)\delta(\vec{r}_{\Lambda}-\vec{r}_N)\\
&+\frac{1}{2}t_1^\Lambda\left(\vec{k^\prime}^2\delta(\vec{r}_{\Lambda}-\vec{r}_N)
+\delta(\vec{r}_{\Lambda}-\vec{r}_N)\vec{k}^2\right)\\
&+t_2^\Lambda\vec{k'}\delta(\vec{r}_\Lambda-\vec{r}_N)\cdot\vec{k}
+iW_0^\Lambda\vec{k'}\delta(\vec{r}_\Lambda-\vec{r}_N)\cdot(\vec{\sigma}
\times\vec{k})
\end{split}
\label{LN}
\end{equation}
and 
\begin{equation}
v_{\Lambda NN}(\vec{r}_\Lambda,\vec{r}_{N_1},\vec{r}_{N_2})=
t_3^\Lambda\delta(\vec{r}_\Lambda-\vec{r}_{N_1})
\delta(\vec{r}_{\Lambda}-\vec{r}_{N_2}),
\label{LNN}
\end{equation}
respectively.
Here, $P_\sigma$ is the spin exchange operator, and 
the operator $\vec{k'}=-(\vec{\nabla}_1-\vec{\nabla}_2)/2i$ acts on the left
hand side while $\vec{k}=(\vec{\nabla}_1-\vec{\nabla}_2)/2i$ acts on 
the right hand side.
See Refs. \cite{Minato2012,Rayet,Lanskoy1998} for the 
details of the Hartree-Fock method for hypernuclei. 

The operator for the electric dipole (E1) response for single-$\Lambda$ 
hypernuclei is given by
\begin{equation}
\hat{F}_\mu
=e\,\sum_{i\in p}(r_iY_{1\mu}(\hat{r}_i)-RY_{1\mu}(\vec{R})), 
\label{e1ope0}
\end{equation}
where 
\begin{equation}
\vec{R}=\frac{1}{M}\,\left(
  m\sum_{i\in n,p} \vec{r}_i
+ m_\Lambda\vec{r}_\Lambda
\right)
\end{equation}
is the center of mass of the hypernucleus.
Here, $M \equiv m(Z+N)+m_\Lambda$ is the total mass of the hypernuclei, 
$m=(m_p+m_n)/2=938.92$ MeV/$c^2$ and $m_\Lambda=1115.68$ MeV/$c^2$
being the mass of nucleon and $\Lambda$ hyperon, respectively.
$N$ and $Z$ are the number of neutron and proton, respectively.

We rearrange the E1 operator, Eq. (\ref{e1ope0}), as 
\begin{equation}
\hat{F}_\mu=\hat{F}_\mu^{({\rm core})}+\hat{F}_\mu^{(\Lambda)},
\end{equation}
with 
\begin{equation}
\hat{F}_\mu^{(\rm core)}=
e
\left(\frac{N}{N+Z} \sum_{i\in p}r_iY_{1\mu}(\hat{r_i}) 
     -\frac{Z}{N+Z} \sum_{i\in n}r_iY_{1\mu}(\hat{r_i})\right),
\label{e1opeC}
\end{equation}
and
\begin{equation}
\hat{F}_\mu^{(\Lambda)}=
-e\frac{Zm_\Lambda}{M}
\Big(r_\Lambda Y_{1\mu}(\hat{r_\Lambda})-R_cY_{1\mu}(\hat{R_c})\Big),
\label{e1opeL}
\end{equation}
where
%

%
%
%
%
\begin{equation}
\vec{R}_c=\frac{1}{N+Z}\sum_{i\in n,p} \vec{r}_i
\end{equation}
is the center of mass of the core nucleus.
The operator $\hat{F}_\mu^{(\Lambda)}$ given by Eq. (\ref{e1opeL}) 
is the $E1$ operator which induces 
the dipole motion between the $\Lambda$ particle and the core nucleus, 
while $\hat{F}_\mu^{{\rm (core)}}$, Eq. (\ref{e1opeC}), 
is identical to the usual $E1$ operator, which generates 
an oscillation between protons and neutrons in a normal nucleus.

The soft dipole $\Lambda$ mode is interpreted as 
the vibration for the relative motion between the $\Lambda$ hyperon 
and the core
nucleus, as we have confirmed with RPA  \cite{Minato2012}. 
We thus estimate its energy by using the sum rules for 
the operator $\hat{F}_\mu^{(\Lambda)}$. 
The energy weighted sum rule can then be calculated using the 
Hamiltonian $\hat{H}$ and the wave function of the ground state, 
$|0\rangle$, as 
\begin{eqnarray}
m_1^{\Lambda}&=&\frac{1}{2}\,\left\langle 0\left| 
\left[\hat{F}^{(\Lambda)}_{0},\left[H,\hat{F}^{(\Lambda)}_{0}\right]\right] 
\right|0 \right\rangle, \\
&=&\frac{3e^2}{4\pi}\left(\frac{Zm_\Lambda}{M}\right)^2\Bigg[
\frac{\hbar^2}{2m_\Lambda}+\frac{\hbar^2}{2m_N(N+Z)} \nonumber \\
&&+\frac{1}{4}\left(t_1^\Lambda +t_2^\Lambda \right)
\left(1+\frac{1}{N+Z}\right)^2
\int\rho_\Lambda(\vec{r})\rho_c(\vec{r})
d\vec{r}
\Bigg], \nonumber \\
\label{m1}
\end{eqnarray}
%
%
where 
$\rho_c(\vec{r})\equiv\rho_p(\vec{r})+\rho_n(\vec{r})$ 
and $\rho_\Lambda(\vec{r})$ 
are the density distributions for the core nucleus and for the $\Lambda$ 
particle, respectively. 
On the other hand, the non-energy weighted sum rule for 
the soft dipole $\Lambda$ mode 
is simply given by 
\begin{equation}
\begin{split}
m_0^\Lambda
&=\sum_{\nu} \left| \left\langle \nu \left|
\hat{F}_0^{(\Lambda)}\right|0\right\rangle \right|^2 
=\left\langle 0\left| \; 
\left(\hat{F}_0^{(\Lambda)}\right)^2 \right|0 \right\rangle\\
&\simeq \left(e\frac{Zm_\Lambda}{M}\right)^2 
\Bigg[ \int \rho_\Lambda(r_\Lambda)
\left(r_\Lambda Y_{10}(\hat{\vec{r}}_\Lambda)\right)^2d\vec{r}_\Lambda \\
& +\int\rho_c(\vec{r})\frac{1}{(N+Z)^2}
\left(rY_{10}(\hat{\vec{r}})\right)^2\, d\vec{r} \Bigg]\\
&=\frac{e^2}{4\pi}\left(\frac{Zm_\Lambda}{M}\right)^2 
\Big( \langle r^2 \rangle_\Lambda +\frac{1}{(N+Z)^2}
\langle r^2 \rangle_c \Big),
\end{split}
\label{m0}
\end{equation}
where $\sqrt{\langle r^2 \rangle_\Lambda}$ and $\sqrt{\langle r^2 \rangle_c}$
are the root mean square (rms) radius of 
the $\Lambda$ hyperon and the core nucleus, respectively. 
Here, we have assumed the perfect decouping between the soft dipole $\Lambda$ 
mode and the giant dipole resonance\cite{SH90}. 
The excitation energy for the soft dipole $\Lambda$ mode is then 
estimated as $E_{sd\Lambda}=m_1^\Lambda/m_0^\Lambda$. 



Let us now numerically evaluate the energy of the soft dipole $\Lambda$ mode 
and discuss its mass number dependence. To this end, we study 
the following nine hypernuclei: $^{16}_{~\Lambda}$O, 
$^{32}_{~\Lambda}$S, $^{40}_{~\Lambda}$Ca, $^{51}_{~\Lambda}$V, 
$^{64}_{~\Lambda}$Ni, $^{89}_{~\Lambda}$Y, $^{120}_{~~\Lambda}$Sn, 
$^{139}_{~~\Lambda}$La and $^{208}_{~~\Lambda}$Pb.
We employ the $\Lambda N$ forces constructed in Ref. \cite{Yamamoto1988}, 
while we use the SkM$^*$ set \cite{SkM*} for the $NN$ interaction. 
There are 6 parameter sets for the $\Lambda N$ interaction, 
No. 1-6, in which the sets No. 1, 2, 5, and 6 
include the 3-body term proportional to $t_3^\Lambda$ 
while No. 3 and No. 4 do not. 
The 3-body term, which works repulsively in general, 
significantly affects the $\Lambda$ binding energy. 
For instance, one finds in Tab. I of Ref. \cite{Yamamoto1988} that the 
No. 1, 2, 5, and 6 sets provide a relatively 
shallow potential depth in nuclear matter $D_\Lambda$,
among which the No. 2 set binds a $\Lambda$ hyperon most weakly 
($D_\Lambda=26.5$ MeV).
On the other hand, the No. 3 and 4 sets without the 3-body term lead to 
a deeper potential depth ($D_\Lambda=32.6$ and $34.6$ MeV, respectively).

In solving the Hartree-Fock equations, we assume spherical symmetry 
for all the nuclei considered in this paper. 
For odd-mass nuclei, we employ the 
filling-approximation \cite{PMR08,BBN09}, with which 
the last occupied orbit is filled only partially so as to reproduce 
the particle number of the whole system. 
For the $\Lambda$ hyperon, we assume that it occupies the 
$1s_{1/2}$ state with the occupation probability of one half.

\begin{figure}
\includegraphics[width=\linewidth]{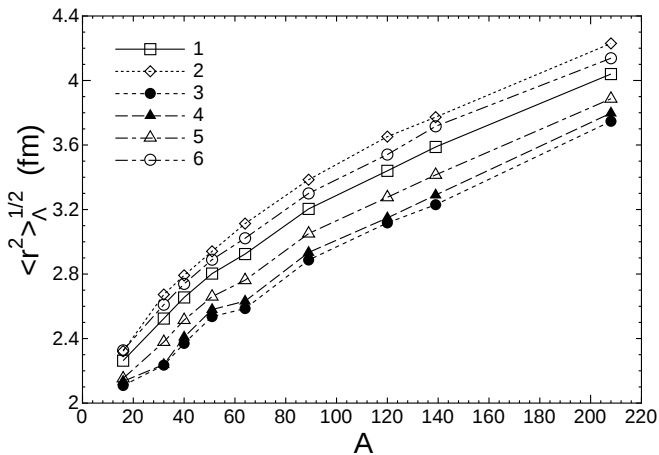}
\caption{The root mean square radius 
$\sqrt{\langle r^2 \rangle_\Lambda}$ for the $\Lambda$ hyperon in 
single-$\Lambda$ hypernuclei 
as a function of the mass number $A$, obtained with the Skyrme-Hartree-Fock 
calculations with six different $\Lambda N$ interactions.}
\label{rmsr}
\end{figure}

We first discuss the rms radius for the $\Lambda$ hyperon, 
$\sqrt{\langle r^2 \rangle_\Lambda}$,  
to which the the non-energy weighted sum rule, $m_0^\Lambda$, is strongly 
related. 
Figure \ref{rmsr} shows 
$\sqrt{\langle r^2 \rangle_\Lambda}$ as a function of $A$, where the results 
with those $\Lambda N$ interactions with the 3-body term (that is, 
the No. 1,2,5, and 6 sets) are denoted by open 
symbols while those without it (that is, No. 3 and 4 sets) 
are by filled symbols. 
Since the radius of the core nucleus 
increases with its mass number, 
it is natural that the rms radius for the $\Lambda$ hyperon, 
$\sqrt{\langle r^2 \rangle_\Lambda}$, 
also increases as a function of the mass 
number, $A$. 
One can see that the calculated radii show a clear dependence 
on the $\Lambda N$ interaction adopted. 
In particular, the interactions with the 3-body 
term ({\it i.e.,} No. 1, 2, 5 and 6) yield a larger radius than 
those without it ({\it i.e.,} No. 3 and 4).
This is because the $\Lambda$ binding energy calculated 
with the No. 1, 2, 5 and 6 sets
is smaller than that with the No. 3 and 4 sets.
As a consequence, the density distribution of the $\Lambda$ hyperon tends to 
expend outward. 
Notice that the No. 2 interaction having 
the largest $t_3^\Lambda$ ($=3000$ MeV$\cdot$fm$^{6}$) leads to 
the largest radii.

\begin{figure}
\includegraphics[width=\linewidth]{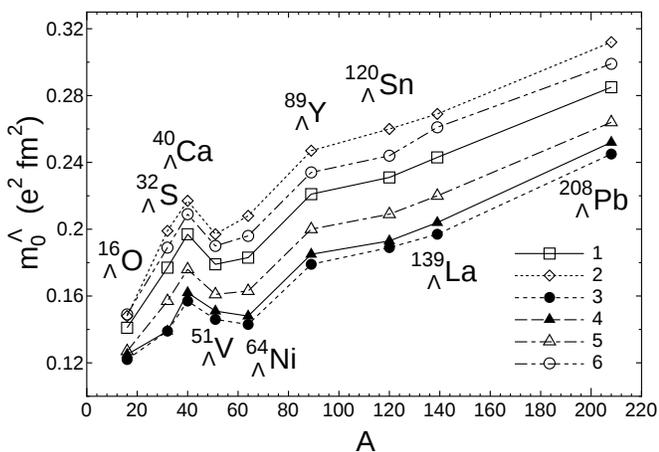}
\caption{The non-energy weighted sum rule 
for the soft dipole $\Lambda$ mode 
as a function of mass number $A$
calculated with six different $\Lambda N$ interactions.}
\label{figm0}
\end{figure}

Figure \ref{figm0} shows the non-energy weighted sum rule, $m_0^\Lambda$. 
On average, it increases with the mass number $A$, similarly to the rms 
radius shown in Fig. \ref{rmsr}. 
It is also similar to the rms radii that the 
the interactions with the 3-body term yield 
relatively larger values for $m_0^\Lambda$. 
However, in contrast to the rms radii, 
which increases monotonically as a function of $A$, the 
non-energy weighted sum rule $m_0^\Lambda$ 
shows a non-monotonic behavior at 
$^{51}_{\;\Lambda}$V and $^{64}_{\;\Lambda}$Ni. 
This can be attributed to the fact that the 
factor $(Zm_\Lambda/M)^2$ in Eq. \eqref{m0} 
decreases from $^{40}_{\;\Lambda}$Ca 
to $^{51}_{\;\Lambda}$V, and then to $^{64}_{\;\Lambda}$Ni due to the deviation 
from the $N=Z$ line (to be more precise, it is the $N-1=Z$ line). 
That is, the value of 
$(Zm_\Lambda/M)^2$ is 0.591, 0.534, and 0.518 for 
$^{40}_{\;\Lambda}$Ca, $^{51}_{\;\Lambda}$V, and $^{64}_{\;\Lambda}$Ni, 
respectively. 

\begin{figure}
\includegraphics[width=\linewidth]{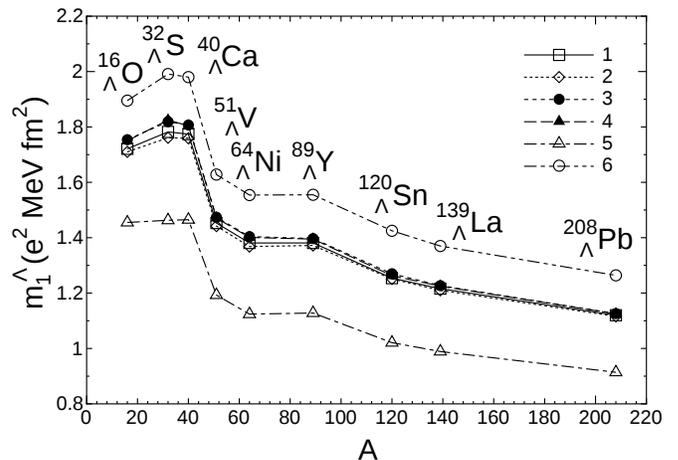}
\caption{The energy weighted sum rule for the soft dipole $\Lambda$ mode 
in single-$\Lambda$ hypernuclei 
as a function of mass number $A$.}
\label{figm1}
\end{figure}


The energy weighted sum rule $m_1^\Lambda$ is shown in Fig. \ref{figm1}. 
The No. 1-4 interactions show almost identical lines, 
as the coefficient of the third term in Eq. \eqref{m1},
$a_1=(t_1^\Lambda+t_2^\Lambda)/4$, is identical for these sets (that is, 
$a_1=26.3$ MeV$\cdot$fm$^5$) \cite{Yamamoto1988}. 
On the other hand, $a_1=0$ and $45.0$ MeV$\cdot$fm$^5$ for the 
No. 5 and 6 parameter sets, 
respectively,
and these parameter sets yield the lowest and the largest value 
for $m_1^\Lambda$, respectively. 
A large decrease of $m_1$ from $^{40}_{~\Lambda}$Ca to $^{51}_{~\Lambda}$V 
is again attributed to 
the decrease of the factor $(Zm_\Lambda/M)^2$ in Eq. \eqref{m1} 
due to the deviation from the $N=Z$ line. 


\begin{figure}
\includegraphics[width=\linewidth]{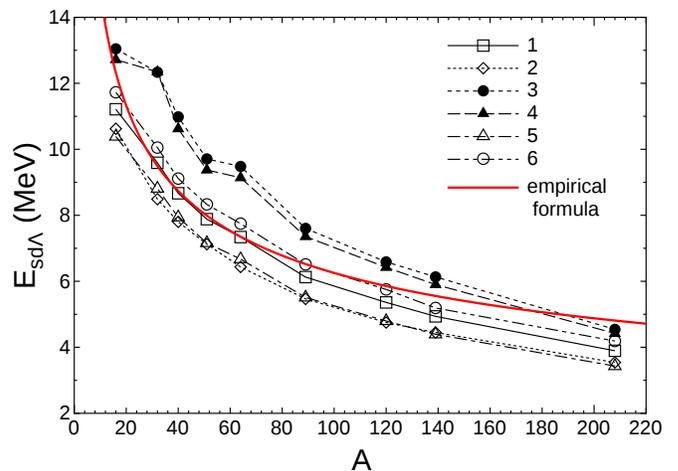}
\caption{(Color online) 
The excitation energy of the soft $\Lambda$ dipole mode, $E_{sd\Lambda}$, 
obtained with the sum rule approach as $E_{sd\Lambda}=m_1^\Lambda/m_0^\Lambda$. 
The thick solid line indicates the results of the empirical formula given by 
Eq. \eqref{esd}. }
\label{figm10}
\end{figure}

The energy of the soft dipole $\Lambda$ mode, $E_{sd\Lambda}$, 
calculated with $m_0^\Lambda$ and $m_1^\Lambda$, is shown in 
Fig. \ref{figm10}. 
Because of the minor contribution of the second term in Eq. \eqref{m0} 
as well as a cancellation of the factor $(Z/M)^2$ between $m_1^\Lambda$ 
and $m_0^\Lambda$, 
the energy $E_{sd\Lambda}$ is almost inversely proportional to 
$\langle r^2 \rangle_\Lambda$. 
As a result, the No. 3 and 4 sets for the $\Lambda N$ interaction, 
giving small root mean square radii, show large values for $E_{sd\Lambda}$, 
while the sets No. 1, 2, 5 and 6 provide small values of $E_{sd\Lambda}$.
For all the $\Lambda N$ interactions, the energy of the soft dipole $\Lambda$ 
mode, $E_{sd\Lambda}$, decreases with the 
mass number $A$ as expected. 
Assuming that the soft dipole $\Lambda$ mode is close to a single particle
excitation of the $\Lambda$ hyperon 
in a harmonic oscillator potential, the mass number dependence of 
$E_{sd\Lambda}$ may be parametrized as 
\begin{equation}
E_{sd\Lambda}=\alpha A^{-1/3} + \beta A^{-2/3}. 
\label{esd}
\end{equation}
By performing the least square fit to 
all the data points obtained with the six different 
$\Lambda N$ interactions, 
we obtain the coefficients of the empirical formula as 
$\alpha$=26.6 MeV and $\beta=11.2$ MeV. 
The energy obtained with this empirical formula is 
shown in Fig. \ref{figm10} by the thick solid line. 
It well reproduces the average behavior of the excitation energy 
of the soft dipole $\Lambda$ mode, even though 
it somewhat overestimates the energy for the $^{208}_{~~\Lambda}$Pb nucleus. 


Shell model calculations with a meson-exchange YN interaction 
have been carried out for low-lying excited states in 
$^{16}_{~\Lambda}$O \cite{Tzeng2000} and $^{40}_{~\Lambda}$Ca 
\cite{Tzeng2002}. 
The soft dipole $\Lambda$ mode appears in these calculations 
at around 10 MeV and 8-9 MeV for 
$^{16}_\Lambda$O and $^{40}_\Lambda$Ca, respectively, although the 
nature of the soft dipole $\Lambda$ mode was not discussed in 
Refs. \cite{Tzeng2000,Tzeng2002}. 
These results are in a reasonable agreement with 
our results obtained with the parameter sets which include the 
three-body $\Lambda NN$ interaction, that is, No. 1,2,5, and 6. 
%

In summary, we have discussed the soft dipole $\Lambda$ mode in 
single-$\Lambda$ hypernuclei using the sum rule approach. 
We have found that the non-energy weighted sum rule, $m_0^\Lambda$, 
for the soft dipole $\Lambda$ mode depends significantly on the 
$\Lambda N$ interaction.
In particular, the $\Lambda N$ interactions with the 3-body term leads to 
smaller values of $m_0^\Lambda$ as compared to those without the 3-body term. 
On the other hand, the energy weighted sum rule, $m_1^\Lambda$, 
has a strong dependence on 
the momentum-dependent terms in the $\Lambda N$ interaction. 
We have argued that the excitation energy of the 
soft dipole $\Lambda$ mode is almost inversely 
proportional to the mean square radius 
and decreases with mass number.
We have derived an empirical formula for the excitation 
energy, that scales as $E_{sd\Lambda}=26.2A^{-1/3}+11.2A^{-2/3}$ MeV. 
%

Our calculations indicate that the soft dipole $\Lambda$ mode appears 
at around 10 MeV in $^{16}_\Lambda$O, which is in agreement with 
a shell model calculation with a meson exchange YN interaction. 
Although the core nucleus, $^{15}$O, is unstable and all the levels are 
not known exactly, 
the strength of the soft dipole $\Lambda$ mode of $^{16}_{\Lambda}$O 
is strong, and it could be distinguished experimentally from the other levels 
associated with the core excitation. In heavier hypernuclei, 
the energy of the soft dipole mode decreases, and for {\it e.g.,} $^{208}_{~~\Lambda}$Pb it appears at around 
4 MeV. In this energy region, there are only 40 discrete levels 
observed in the core nucleus, $^{207}$Pb \cite{nudat}. 
One can thus have a hope to experimentally 
identify the soft dipole $\Lambda$ mode in single-$\Lambda$ hypernuclei. 
It would be extremely interesting if such measurement could be 
realized in some future. 

\medskip

This work was supported by 
JSPS KAKENHI Grant Numbers 22540262 and 25105503.

\end{document}